\documentclass[a4paper,12pt]{article}

\usepackage{graphics, epsfig}

\begin{document}

\author{E. M. O'Shea
\thanks{E-mail : emer.oshea@ucd.ie}\\
\small Mathematical Physics Department\\ \small  National
University of Ireland Dublin, Belfield, Dublin 4, Ireland}

\title{Properties of Information Carrying Waves in Cosmology}

\date{}

\maketitle

\begin{abstract}

Recently we studied the effects of information carrying waves
propagating through isotropic cosmologies. By information carrying
we mean that the waves have an arbitrary dependence on a function.
We found that the waves introduce shear and anisotropic stress
into the universe. We then constructed explicit examples of pure
gravity wave perturbations for which the presence of this
anisotropic stress is essential and the null hypersurfaces playing
the role of the histories of the wave--fronts in the background
space--time are shear--free. Motivated by this result we now prove
that these two properties are true \emph{for all information
carrying waves} in isotropic cosmologies.

\end{abstract}

\vskip 2truepc\noindent

PACS number(s): 04.30.Nk

\thispagestyle{empty}

\newpage

\section{Introduction}\indent

In a recent paper \cite{isotropic} the propagation of shear--free
gravitational waves through isotropic universes was studied using
the gauge--invariant and covariant formalism of Ellis--Bruni
\cite{EB}. In this approach the waves are represented as a
perturbation of the Robertson--Walker space--time. The
gravitational waves studied also had the property that they could
carry arbitrary information. By this we mean that the Ellis--Bruni
gauge--invariant variables had an arbitrary dependence on a
function.

It was found that the consistency of the partial differential
equations governing the perturbations requires that all
gauge--invariant perturbed quantities must vanish except for the
perturbed shear of the matter world--lines and the anisotropic
stress of the matter distribution. A wave equation for the
perturbed shear was derived. In addition the perturbed shear was
shown to satisfy a propagation equation along the null rays
associated with the gravitational waves (see eqs. (\ref{2.7}) and
(\ref{2.8}) below).

For pure gravitational radiation (i.e. the perturbed Weyl tensor
is type $N$ in the Petrov classification) explicit examples were
constructed for which it was shown (using the wave equation and
the propagation equation) that the presence of the anisotropic
stress is essential and the hypersurfaces playing the role of the
histories of the wave--fronts are shear--free. We now prove that
these two properties are true in general and are not in any way
dependent on the examples constructed.

The paper is organized as follows: The notation used and some
useful equations are given in the Appendix and referred to where
appropriate. In Section 2 we outline the results obtained in
\cite{isotropic} which we use in the paper. The first of our
results is also given in this section. The elements of the
Newman--Penrose formalism necessary for this problem are described
in Section 3 and the main result is derived in Section 4. The
paper ends with a summary of the results of Sections 3 and 4.

\setcounter{equation}{0}

\section{Equations Governing Gravitational Waves}\indent

\setcounter{equation}{0} Throughout this paper we shall use the
notation and sign conventions of \cite{Ellis}. For convenience
these have been briefly outlined in the Appendix. We are concerned
with a four dimensional space--time manifold with metric tensor
$g_{ij}$ in a local coordinate system ${x^i}$ and a preferred
congruence of world--lines tangent to the unit vector field $u^i$
(with $u^i\,u_i=-1$). The decompositions of the symmetric
energy--momentum--stress tensor $T_{ij}$ and the Weyl tensor with
respect to this 4--velocity field, along with definitions of the
kinematical quantities $\sigma_{ij}$, $\omega_{ij}$, $\theta$, can
be found in the Appendix.

In a recent paper \cite{isotropic} we used the gauge--invariant
and covariant approach of Ellis--Bruni \cite{EB} to construct
gravitational wave perturbations of Robertson--Walker
space--times. The background metric tensor is thus the
Robertson--Walker metric, the background energy--momentum--stress
tensor is (\ref{A9}) specialized to a perfect fluid (i.e. with
$q^i=0=\pi^{ij}$) with fluid 4--velocity $u^i$ and the background
Weyl tensor vanishes. The Ellis--Bruni approach involves working
in a general local coordinate system with gauge--invariant small
quantities rather than small perturbations of the background
metric. These quantities have the property that they vanish in the
background space--time. For isotropic space--times the
Ellis--Bruni variables are $\sigma_{ij}$, $\dot{u}^i$,
$\omega_{ij}$, $X_{i}=h^j_i\mu_{,\,j}$, $Y_i=h^j_ip_{,\,j}$,
$Z_i=h^j_i\theta_{,\,j}$, $\pi_{ij}$, $q_i$, $E_{ij}$ and $H_{ij}$
(see the Appendix for definitions of these variables). We found
that it is tensor quantities that describe gravitational wave
perturbations and thus we can set all Ellis--Bruni variables equal
to zero except for $\pi_{ij}$, $\sigma_{ij}$, $E_{ij}$ and
$H_{ij}$. The equations satisfied by these variables are obtained
by projections, in the direction of and orthogonal to the
4--velocity $u^i$, of the equations of motion and the energy
conservation equation contained in $T^{ij}{}_{;\,j}=0$, the Ricci
identities and the Bianchi identities (see Appendix). These
projections give rise to a lengthy list of equations which can be
found in \cite{isotropic}.

We specialized to a particular class of gravitational waves,
namely those which can be viewed as carrying arbitrary
information. To incorporate this into the Ellis--Bruni set up we
required that the gauge--invariant small quantities have an
arbitrary dependence on a function. Specifically we assume that
\begin{equation}\label{2.1} \pi_{ij}=\Pi_{ij}\,F(\phi )\ , \qquad
\sigma_{ij}=s_{ij}F(\phi )\ ,\end{equation} where $F$ is an
arbitrary real--valued function of its argument $\phi(x^i)$. This
idea of introducing arbitrary functions into solutions of
Einstein's equations describing gravitational waves goes back to
work by Trautman \cite{trautman} and this form for the
Ellis--Bruni variables was first used by Hogan and Ellis
\cite{H+E}. Using the projections of the Ricci identities with
$\omega_{ij}=0=\dot{u}^i$ we can write

\begin{equation}\label{2.2}
E_{ij}=\frac{1}{2}\,\pi_{ij}+\frac{2}{3}\,\sigma^2h_{ij}-
\frac{2}{3}\,\theta\,\sigma_{ij}-\sigma_{ik}\,
\sigma^k{}_j-h^k_i\,h^l_j\,\dot{\sigma}_{kl}\ ,\end{equation}and
\begin{equation}\label{2.3}
H_{ij}=-h^k_i\,h^l_j\,\sigma_{(\,k}{}^{p\,;\,q}\,\eta_{l)rpq}\,u^r\
.\end{equation}Thus these variables are derived from $\sigma_{ij}$
and $\pi_{ij}$ and it is not necessary to make any extra
assumptions about their dependence on $F(\phi)$. We note that
$\Pi_{ij}$, $s_{ij}$ are orthogonal to $u^i$ and trace--free with
respect to the background metric $g_{ij}$.

Substituting (\ref{2.1}) into the linearized versions of the
equations satisfied by these variables we find that $s^{ij}$ and
$\Pi^{ij}$ are orthogonal to the gradient of $\phi$ and divergent
free i.e. we have \begin{equation}\label{2.4}s^{ij}\phi_{,\,j}=0\
, \qquad \Pi^{ij}\phi_{,\,j}=0\ ,
\end{equation} and
\begin{equation}\label{2.5}s^{ij}{}_{;\,j}=0\ , \qquad \Pi^{ij}{}_{;\,j}=0\ .
\end{equation} Also, with $\dot{\phi}:=\phi_{,i}u^i\neq0$, consistency of the
equations requires that
\begin{equation}\label{2.6}g^{ij}\,\phi_{,\,i}\,\phi_{,\,j}=0\ ,
\end{equation} where $g_{ij}$ is the background metric. This means that the
hypersurfaces $\phi(x^i)=\rm{const.}$ in the background isotropic
space--time playing the role of the histories of the wave--fronts
must be null. The following wave equation for $s^{ij}$,
\begin{equation}\label{2.7}s^{ij;k}{}_{;k}-\frac{2}{3}\,\theta\,{\dot s}^{ij}-\left
(\frac{1}{3}\,\dot\theta +\frac{4}{9}\,\theta ^2\right
)\,s^{ij}+(p-\frac{1}{3}\,\mu_0 )\, s^{ij}=-\dot\Pi
^{ij}-\frac{2}{3}\,\theta\,\Pi ^{ij}\ ,
\end{equation} and propagation equation for $s_{ij}$ along the null geodesics
tangent to $\phi^{,\,i}$,
\begin{equation}\label{2.8}s_{ij;\,k}\,\phi^{,\,k}+\left(\frac{1}{2}\,\phi^{,\,k}{}_{;\,k}-\frac{1}{3}\,\theta\,\dot{\phi}\right)s_{ij}=-\frac{1}{2}\,\dot{\phi}\,
\Pi_{ij}\ ,
\end{equation} are consequences of the Ricci identities. We have used $\mu_0$
here to denote the energy density of the cosmic fluid to avoid
confusion with the spin coefficient $\mu$ used in the next
section. The internal consistencies of these equations were
checked in \cite{isotropic}. Substituting (\ref{2.1}) into
(\ref{2.2}) and (\ref{2.3}) we find that the electric and magnetic
parts of the Weyl tensor are now given by
\begin{equation}\label{2.9}
E_{ij}=\left(\frac{1}{2}\,\Pi_{ij}-\dot{s}_{ij}-\frac{2}{3}\,\theta\,s_{ij}\right)\,F-\dot{\phi}\,s_{ij}\,F^{'}
\ ,
\end{equation} and \begin{equation}\label{2.10}
H_{ij}=-s_{(i}{}^{l\,;\,m}\,\eta_{j\,)klm}\,u^k\,F-s_{(i}{}^l\,\eta_{j)klm}\,u^k\,\phi^{\,,\,m}F^{'}\
,
\end{equation} where $F^{'}=\frac{\partial F}{\partial \phi}$.

We are interested here in pure gravitational wave perturbations
i.e. perturbations which have pure type $N$ perturbed Weyl tensor
in the Petrov classification. It is easily checked using Eq.
(\ref{2.4}) that the $F^{'}$--parts of $E_{ij}$ and $H_{ij}$ above
are automatically type $N$ with degenerate principal null
direction $\phi^{,\,i}$. However the $F$--parts of $E_{ij}$ and
$H_{ij}$ are not in general type $N$ and so to describe pure
gravitational wave perturbations we require (see \cite{isotropic})
that
\begin{equation}\label{2.11}
\dot{s}^{ij}\,\phi_{,\,j}=0\ , \qquad
s^{ij}\,\phi_{,\,i}{}^{;\,k}-s^{ik}\,\phi_{,\,i}{}^{;\,j}=0\ .
\end{equation}

Making use of the following null tetrad,
$k_i=-\dot{\phi}^{-1}\,\phi_{,\,i}$, $l_i=u_i-\frac{1}{2}k_i$, and
$m_i$, ${\bar{m}_i}$ a complex covariant vector field and its
complex conjugate chosen so that they are null
($m^i\,m_i=0={\bar{m}}^i\,{\bar{m}}_i$), are orthogonal to $k^i$
and $l^i$ and satisfy $m^i\,{\bar{m}}_i=1$ we find that (see
\cite{isotropic}) the conditions (\ref{2.11}) are equivalent to

\begin{equation}\label{2.12}
s\,\phi_{,\,i;\,j}\,{\bar{m}}^i\,l^j=0
\end{equation} and \begin{equation}\label{2.13}
{\bar{s}}\,\phi_{,\,i;\,j}m^i\,m^j=s\,\phi_{,\,i;\,j}{\bar{m}}^i\,{\bar{m}}^j\
.
\end{equation} In terms of the tetrad we can write $s^{ij}$ as (because
$s^{ij}u_j=0$, $s^i{}_i=0$, $s^{ij}\,\phi_{,\,j}=0$ and so
$s^{ij}k_j=0$ and $s^{ij}l_j=0$)
\begin{equation}\label{2.14}
s^{ij}={\bar{s}}m^im^j+s{\bar{m}}^i{\bar{m}}^j\ ,
\end{equation}
where $|s|^2=\frac{1}{2}s^{ij}s_{ij}$. For future use we point out
here that $\Pi^{ij}$ can be expressed on the tetrad in a similar
way, namely,
\begin{equation}\label{2.15}
\Pi^{ij}={\bar{\Pi}}m^im^j+\Pi{\bar{m}}^i{\bar{m}}^j\ .
\end{equation}
We emphasize that we are using $\Pi$ here for the complex
component of $\Pi^{ab}$ to avoid confusion with the spin
coefficient $\pi$ appearing in the next section. Taking the
divergence of the first of Eqs. (\ref{2.4}) and substituting for
$s^{ij}$ from (\ref{2.14}) yields (since $s^{ij}{}_{;\,j}=0$)
\begin{equation}\label{2.16}
({\bar{s}}\,m^im^j+s\,{\bar{m}}^i{\bar{m}}^j)\phi_{,\,j;\,i}=0\ .
\end{equation} Comparing this with Eq. (\ref{2.13}) it is clear that for
consistency (with $s\neq0$) we must have
\begin{equation}\label{2.17}
\phi_{,\,i;\,j}m^i\,m^j=0\ .
\end{equation} Thus \emph{the hypersurfaces} $\phi(x^i)=\rm{const}.$ which play
the role of the histories of the wave--fronts \emph{must be
shear--free}. With $s\neq0$ (\ref{2.12}) becomes
\begin{equation}\label{2.18}
\phi_{,\,i;\,j}{\bar{m}}^i\,l^j=0\ .
\end{equation} This condition corresponds to the vanishing of the spin
coefficients $\tau$, $\nu$ which we shall make use of in the next
section.

\setcounter{equation}{0}

\section{The Newman--Penrose Formalism}\indent

For the remainder of the paper we seek to demonstrate that the
presence of anisotropic stress in the perturbed space--time is
essential for the existence of pure gravitational waves. We find
the most effective way to do this is to make use of the
Newman--Penrose Formalism. A comprehensive description of this
formalism can be found in \cite{chand}. Specifically we choose the
tetrad described after Eq. (\ref{2.11}) and we label the vectors
as follows:

\begin{equation}\label{3.1}
\begin{array}{cc}
D=k^i\,\partial_i=e^i_1\,\partial_i\ , &
\Delta=l^i\,\partial_i=e^i_2\,\partial_i\ , \\
\delta=m^i\,\partial_i=e^i_3\,\partial_i\ , &
\delta^*={\bar{m}}^i\,\partial_i=e^i_4\,\partial_i\ .
\end{array}
\end{equation} The spin coefficients (Ricci rotation coefficients) are denoted by
$\gamma_{abc}=-\gamma_{bac}$. It follows from Eq. (\ref{2.17})
that $k^i$ is shear--free (i.e. $k_{i;j}m^im^j=0$) and a simple
calculation using the fact that we can write
$\dot{\phi}_{,b}=(\phi_{,b})^.+\frac{1}{3}\theta h^c_b\phi_{,c}$
shows that $k^i$ is also geodesic with
$k^i{}_{;j}k^j=\frac{1}{3}\theta k^i$. As a consequence of these
properties the spin coefficients $\kappa, \sigma, \lambda$ and
$\pi$ vanish. We again point out that $\pi$ here is not to be
confused with the complex component $\Pi$ of $\Pi^{ij}$. As
previously mentioned after Eq. (\ref{2.18}) the spin coefficients
$\nu$ and $\tau$ also vanish. Therefore, for the problem at hand
the only non--zero background spin coefficients are $\rho, \mu,
\gamma, \epsilon, \alpha, \beta$ with

\begin{eqnarray}\label{3.2}
\rho&=& -\frac{1}{2}\dot{\phi}^{-1}\phi^{,a}{}_{;a}\ ,\\ \mu&=&
\frac{1}{2}\rho-\frac{1}{3}\theta\ , \\ \epsilon&=&
\frac{1}{2}(\gamma_{211}+\gamma_{341})\ ,\\ \gamma&=&
\frac{1}{2}(\gamma_{212}+\gamma_{342})\ , \\ \alpha&=&
\frac{1}{2}(\gamma_{214}+\gamma_{344})\ , \\ \beta&=&
\frac{1}{2}(\gamma_{213}+\gamma_{343})\ .
\end{eqnarray}

Using $\gamma_{abc}=-\gamma_{bac}$ and noting that the
4--acceleration vanishes in the background we find that

\begin{equation}\label{3.8}
\gamma+\gamma^*=\frac{1}{6}\theta\ , \qquad
\epsilon+\epsilon^*=-\frac{1}{3}\theta\ ,
\end{equation}where, following Chandrasekhar, the star denotes complex
conjugation. In addition $\phi$ is a real--valued function and
thus $\rho$ and $\mu$ are real and we have

\begin{equation}\label{3.9}
\rho-\rho^*=0\ , \qquad \mu-\mu^*=0\ .
\end{equation}We will use these facts in the next section when we examine the
wave equation (\ref{2.7}) and the propagation equation (\ref{2.8})
in this formalism.

In the tetrad formalism the Ricci identities (\ref{A7}) read

\begin{equation}\label{3.10}
R_{abcd}=-\gamma_{abc,d}+\gamma_{abd,c}+\gamma_{baf}(\gamma_{c}{}^f{}_d-\gamma_d{}^f{}_c)+\gamma_{fac}\gamma_{b}{}^f{}_d
-\gamma_{fad}\gamma_b{}^f{}_c\ ,
\end{equation}where $R_{abcd}$ are the tetrad components of the Riemann tensor.
An alternative expression for these components is given via the
definition of the tetrad components of the Weyl tensor, namely,

\begin{eqnarray}\label{3.11}
R_{abcd}&=&C_{abcd}+\frac{1}{2}(\eta_{ac}R_{bd}-\eta_{bc}R_{ad}-\eta_{ad}R_{bc}+\eta_{bd}R_{ac})
\nonumber
\\&-&\frac{1}{6}(\eta_{ac}\eta_{bd}-\eta_{ad}\eta_{bc})R\ .
\end{eqnarray}Here $R_{ab}=R_{ij}e^i_ae^j_b$ are the tetrad components of the
Ricci tensor, $R=\mu_0-3p$ is the Ricci scalar and
$\eta_{ab}:=e^i_ae_{ib}=0$ except for
$\eta_{12}=\eta_{21}=-\eta_{34}=-\eta_{43}=-1$. In general we
label tetrad components of tensors using $a,b,c,\ldots$ and
coordinate components using $i,j,k, \ldots$ .Making use of Eqs.
(\ref{A8}) and (\ref{A9}) we find that the tetrad components of
the background perfect fluid can be written in the form:

\begin{equation}\label{3.12}
R_{ab}=(\mu_0+p)u_au_b+\frac{1}{2}(\mu_0-p)\eta_{ab}\ ,
\end{equation}where $u_a=u_ie^i_a$ with $u_i=l_i+\frac{1}{2}k_i$ and thus
$u_a=(-1,-\frac{1}{2},0,0)$. Here $u_a$ are the tetrad components
of the 4--velocity of a fluid particle, $\mu_0$ is the proper
density and $p$ is the isotropic pressure. We shall always assume
that $\mu_0+p\neq0$ so that $R_{ab}$ has a unique time--like
eigenvector. Evaluating the various components of the Riemann
tensor using both Eqs. (\ref{3.10}) and (\ref{3.11}) (we use the
commutation relations involving the operators $D,\Delta, \delta,
\delta^*$ to do this) and then equating the two different
expressions obtained yields the following twelve equations for the
derivatives of the \emph{background} spin coefficients:

\begin{equation}\label{3.13}
D\rho=-\rho^2-(\epsilon+\epsilon^*)-\frac{1}{2}(\mu_0+p)\ ,
\end{equation}
\begin{equation}\label{3.14}
D\alpha-\delta^*\epsilon=-\alpha(\rho+\epsilon^*-2\epsilon)+\beta^*\epsilon\
,
\end{equation}
\begin{equation}\label{3.15}
D\beta-\delta\epsilon=\beta(\epsilon^*-\rho^*)+\epsilon\alpha^*\ ,
\end{equation}
\begin{equation}\label{3.16}
D\gamma-\Delta\epsilon=2\epsilon\gamma+\epsilon\gamma^*+\epsilon^*\gamma-\frac{1}{12}\mu_0-\frac{1}{4}p\
, \end{equation}
\begin{equation}\label{3.17}
D\mu=-\mu\rho^*+\mu(\epsilon+\epsilon^*)+\frac{1}{4}p-\frac{1}{12}\mu_0\
,
\end{equation}
\begin{equation}\label{3.18}
\delta\rho=-\rho(\alpha^*+\beta)\ ,
\end{equation}
\begin{equation}\label{3.19}
\delta\alpha-\delta^*\beta=-\mu\rho-\alpha\alpha^*-\beta\beta^*+2\alpha\beta-\frac{1}{6}\mu_0\
,
\end{equation}
\begin{equation}\label{3.20}
\delta^*\mu=\mu(\alpha+\beta^*)\ ,
\end{equation}
\begin{equation}\label{3.21}
\Delta\mu=\mu^2+\mu(\gamma+\gamma^*)+\frac{1}{8}(\mu_0+p)\ ,
\end{equation}
\begin{equation}\label{3.22}
\delta\gamma-\Delta\beta=2\beta\gamma+\alpha^*\gamma-\mu\beta-\beta\gamma^*\
,
\end{equation}
\begin{equation}\label{3.23}
\Delta\rho=\rho\mu^*-\rho(\gamma+\gamma^*)-\frac{1}{4}p+\frac{1}{12}\mu_0\
,
\end{equation}
\begin{equation}\label{3.24}
\Delta\alpha-\delta^*\gamma=\mu^*\alpha-\beta^*\gamma-\alpha\gamma^*\
.
\end{equation}Due to the fact that so many of the background spin coefficients
are zero, the remaining Ricci identities give no further
information. We note that we have made use of the vanishing of the
background Weyl tensor in this calculation. Also for clarity we
refrain from substituting for the spin coefficients from Eqs.
(\ref{3.2})--(3.7) until the very end.

Under the (class III) rotation

\begin{equation}\label{3.25}
k^j \rightarrow k^j\ ,\qquad  l^j \rightarrow l^j\ , \qquad m^j
\rightarrow e^{i\varphi}m^j\ , \qquad {\bar{m}}^j \rightarrow
e^{-i\varphi}{\bar{m}}^j\ ,
\end{equation}where $A$ and $\varphi$ are two real functions the non--zero spin
coefficients transform as

\begin{eqnarray}\label{3.26}
\rho &\rightarrow& \rho\ , \\ \mu &\rightarrow& \mu\ , \\ \gamma
&\rightarrow& \gamma+\frac{1}{2}i\Delta\varphi \ ,\\ \epsilon
&\rightarrow& \epsilon +\frac{1}{2}iD\varphi\ , \\ \alpha
&\rightarrow&
e^{-i\varphi}\alpha+\frac{1}{2}ie^{-i\varphi}\delta^*\varphi \ ,\\
\beta
&\rightarrow&e^{i\varphi}\beta+\frac{1}{2}ie^{i\varphi}\delta\varphi\
.
\end{eqnarray}From these transformations we see that we can in principle rotate
away $\epsilon-\epsilon^*$, $\gamma-\gamma^*$ and $\alpha-\beta^*$
(that is we can assume $\epsilon$, $\gamma$ are real and
$\alpha=\beta^*$) provided the function $\varphi$ satisfies

\begin{equation}\label{3.32}
D\varphi=i(\epsilon-\epsilon^*)\ , \qquad
\Delta\varphi=i(\gamma-\gamma^*)\ , \qquad
\delta\varphi=i(\alpha-\beta^*)\ .
\end{equation}However before we impose these conditions we need to check that
they are consistent with each other and with all other equations.
Suppose the first of these two conditions hold. Using Eq.
(\ref{3.16}) we find

\begin{equation}\label{3.33}
D\Delta\varphi-\Delta
D\varphi=2i(\epsilon\gamma-\epsilon^*\gamma^*)
\end{equation}and it is straightforward to check that this agrees with one of
the standard commutation relations. Thus we can set
$\epsilon-\epsilon^*=0$ and $\gamma-\gamma^*=0$ without imposing
any extra conditions. If we assume that the third of (\ref{3.32})
holds then Eq. (\ref{3.19}) and the commutation relation involving
$[\delta, \delta^*]$ are consistent only if we have

\begin{equation}\label{3.34}
\mu^*\rho^*+\mu\rho=-\frac{1}{3}\mu_0\ .
\end{equation}Since $\mu, \rho$ are real this simplifies to

\begin{equation}\label{3.35}
2\mu\rho=-\frac{1}{3}\mu_0\ .
\end{equation}Replacing $\mu$ here by (3.3) yields the quadratic equation
\begin{equation}\label{3.36} 3\rho^2-2\theta\rho+\mu_0=0\ .
\end{equation}In order to have $\rho$ real we require

\begin{equation}\label{3.37}
\theta^2\geq3\mu_0\ .
\end{equation}Here $\theta$ is the expansion of the universe with
$\theta=3\dot{R}/R$ where $R(t)$ is the scale factor of the
general Robertson--Walker line--element and the dot indicates
differentiation with respect to $t$. Using the background Einstein
equations $G_{ij}=(\mu_0+p)u_iu_j+pg_{ij}$, where $G_{ij}$ is the
Einstein tensor, we obtain the background Friedmann equation

\begin{equation}\label{3.38}
\theta^2=3\mu_0-\frac{9k}{R(t)^2}\ ,
\end{equation}where $k=0,\pm1$ is the Gaussian curvature of the homogeneous
space--like hypersurfaces $t=\rm{constant}$. Comparing this with
(\ref{3.37}) we see that if we impose (\ref{3.34}) we cannot
continue to study universes with $k=+1$. However in
\cite{isotropic} we constructed examples of information carrying
waves propagating through such universes and so we shall not
insist on having $\alpha-\beta^*=0$. Thus from this point we
assume only that we can set $\gamma-\gamma^*=0$ and
$\epsilon-\epsilon^*=0$ without loss of generality.

We complete this section by listing (in terms of the spin
coefficients) some derivatives of the tetrad vectors which we use
extensively in the next section:

\begin{equation}\label{3.39}
m_{i;j}m^j=(\alpha^*-\beta)m_i\ ,
\end{equation}

\begin{equation}\label{3.40}
{\bar{m}}_{i;j}m^i=(\alpha-\beta^*)m_j+(\beta-\alpha^*){\bar{m}}_j\
,
\end{equation}

\begin{equation}\label{3.41}
m_{i;j}{\bar{m}}^j=-\mu^*k_i+\rho l_i+(\beta^*-\alpha)m_i\ ,
\end{equation}
\begin{equation}\label{3.42}
k_{i;j}l^j=-(\gamma+\gamma^*)k_i\ ,
\end{equation}
\begin{equation}\label{3.43}
m_{i;j}k^j=0\ ,
\end{equation}
\begin{equation}\label{3.44}
m_{i;j}l^j=0\ .
\end{equation}

\section{The Key Result}

\setcounter{equation}{0} We now have all the equations necessary
to derive our result from the wave equation (\ref{2.7}) for
$s^{ij}$ using the propagation equation (\ref{2.8}) and the
divergence--free condition (\ref{2.5}). To begin with we examine
the divergence--free condition. Writing this condition in tetrad
formalism using (\ref{2.14}) and then contracting with $m_i$
yields

\begin{equation}\label{4.1}
\delta^*s=-2(\alpha-\beta^*)s\ .
\end{equation}
Thus we now have an expression for the derivative of $s$ in the
direction of ${\bar{m}}^i$. In terms of the spin coefficients we
can write

\begin{equation}\label{4.2}
\theta=\frac{3}{2}\rho-3\mu
\end{equation}and

\begin{equation}\label{4.3}
\phi_{,i}{}^{;i}=-2\dot{\phi}\rho\ .
\end{equation}Multiplying the propagation equation (\ref{2.8}) by $m^im^j$ and
using these expressions we arrive at (since
$\phi_{,i}=-\dot{\phi}k_i$) the following equation for the
derivative of $s$ in the direction of $k^i$:

\begin{equation}\label{4.4}
Ds=\left(\mu-\frac{3}{2}\rho\right)s+\frac{1}{2}\Pi\ .
\end{equation}Similarly we multiply the wave equation (\ref{2.7}) by $m^im^j$
and this yields

\begin{eqnarray}\label{4.5}
s^{ij;k}{}_{;k}m_im_j&-&\frac{2}{3}\theta\dot{s}^{ij}m_im_j-\left(\frac{1}{3}\dot{\theta}+\frac{4}{9}\theta^2\right)s+
\left(\rho-\frac{1}{3}\mu_0\right)s\nonumber
\\ &=&-\dot{\Pi}^{ij}m_im_j-\frac{2}{3}\theta\Pi\ .
\end{eqnarray}Writing $s^{ij;k}{}_{;k}$ as $g^{kl}s^{ij}{}_{;kl}$ and using Eq.
(\ref{2.14}) we can (on substitution from Eqs.
(\ref{3.39})--(\ref{3.44})) write the first term of this wave
equation as

\begin{eqnarray}\label{4.6}
s^{ij;k}{}_{;k}m_im_j&=&2[\delta(\delta^*s)-\Delta(Ds)]-2(\gamma+\gamma^*-\mu)Ds+2s\delta(\alpha-\beta^*)
\nonumber \\ &+&2(\beta-\alpha^*)\delta^*s+4(\alpha-\beta^*)\delta
s+2s\delta^*(\beta-\alpha^*) \nonumber \\ &-&2\rho^*\Delta
s+8(\beta-\alpha^*)(\alpha-\beta^*)s-2(\mu^*\rho^*+\mu\rho)s\ .
\end{eqnarray}Replacing $Ds$ here by (\ref{4.4}), $\delta^*s$ by (\ref{4.1}) and
using Eq. (\ref{3.19}) to write $\delta^*\alpha-\delta\beta$ (and
its complex conjugate) in terms of undifferentiated spin
coefficients this equation simplifies to

\begin{eqnarray}\label{4.7}
s^{ij;k}{}_{;k}m_im_j&=&\left[-2\Delta\mu+3\Delta\rho-2\left(\frac{1}{6}\theta-\mu\right)\left(\mu-\frac{3}{2}\rho\right)\right]s
+\frac{2}{3}\theta\Delta s \nonumber \\
&+&\frac{2}{3}\mu_0s-\left(\frac{1}{6}\theta-\mu\right)\Pi-\Delta\Pi\
.
\end{eqnarray}We have also used Eqs. (\ref{3.8}), (\ref{3.9}) and (\ref{4.2}) to
write the equation in this form.

We now turn our attention to the second term of the wave equation
(\ref{4.5}). As a consequence of (\ref{2.14}) we have

\begin{equation}\label{4.8}
\dot{s}^{ij}m_im_j=\dot{s}+2s\dot{{\bar{m}}}^im_i\ ,
\end{equation}where the dot indicates covariant differentiation in the direction
of the 4--velocity $u^i$. Writing out this derivative explicitly
(i.e. write $\dot{s}=s_{;i}u^i$ ) and noting that
$u^i=l^i+\frac{1}{2}k^i$ the above equation becomes (on account of
$\gamma-\gamma^*=\epsilon-\epsilon^*=0$)

\begin{equation}\label{4.9}
\dot{s}^{ij}m_im_j=\Delta s+\frac{1}{2}Ds\ .
\end{equation}The anisotropic stress $\Pi^{ij}$ satisfies the same equations as
$s^{ij}$ (c.f. (\ref{2.4}) and (\ref{2.5}) and thus we also have

\begin{equation}\label{4.10}
\dot{\Pi}^{ij}m_im_j=\Delta \Pi+\frac{1}{2}D\Pi\ .
\end{equation}Now reconstructing the wave equation (\ref{4.5}) using Eqs.
(\ref{4.7}), (\ref{4.9}) and (\ref{4.10}) yields

\begin{eqnarray}\label{4.11}
&
&\left[3\Delta\rho-2\Delta\mu-2\left(\frac{1}{6}\theta-\mu\right)\left(\mu-\frac{3}{2}\rho\right)
-\frac{1}{3}\theta\left(\mu-\frac{3}{2}\rho\right)\right]s\nonumber
\\ &+&
\left(p+\frac{1}{3}\mu_0\right)s-\left(\frac{1}{3}\dot{\theta}+\frac{4}{9}\theta^2\right)s=-\frac{1}{2}D\Pi-\frac{1}{3}\theta\Pi-\mu_0\Pi
\ .
\end{eqnarray}The background Raychaudhuri equation is
\begin{equation}\label{4.12}
\dot{\theta}=-\frac{1}{3}\theta^2-\frac{1}{2}(\mu_0+3p)\ .
\end{equation}Making use of this, the expressions for $\Delta\mu$ and $\Delta
\rho$ given by Eqs. (\ref{3.17}) and (\ref{3.21}) respectively and
Eq. (\ref{4.2}) we find that the wave equation (\ref{4.11})
reduces to the remarkably simple form

\begin{equation}\label{4.13}
(\mu_0+p)s=-D\Pi-\rho\Pi\ .
\end{equation}

The necessity for $\Pi\neq0$ follows from this equation because it
is immediately clear that if the perturbed anisotropic stress
vanishes ($\Leftrightarrow \Pi=0$) then the perturbed shear is
also zero ($\Leftrightarrow s=0$) if $\mu_0+p\neq0$ and we have no
perturbations (i.e. no gravitational waves). We note that the
right hand side of equation (\ref{4.13}) vanishes if $\Pi$
satisfies the differential equation $D\Pi+\rho\Pi=0$. In this case
we must have $s=0$ (since $\mu_0+p\neq0$) and thus from the
propagation equation (\ref{4.4}) we must have $\Pi=0$.

\section{Summary}\indent

We have investigated general properties of some gravity wave
perturbations of isotropic cosmologies. Specifically we have
considered gravitational waves carrying arbitrary information. By
this we mean that the perturbations describing the waves have an
arbitrary dependence on a real--valued function $\phi(x^i)$. Our
main result hinges on two important properties; in order to have
pure gravity waves the hypersurfaces $\phi(x^i)$ must be
\emph{null} and \emph{shear--free}. Now our main result (proved in
Section 4) is that under the physically reasonable assumption that
$\mu_0+p\neq0$ (where $\mu_0$ and $p$ are the proper density and
pressure of the cosmic fluid) the perturbations describing the
gravitational waves \emph{must} be accompanied by the presence of
anisotropic stress in cosmic fluid.

\noindent

\section*{Acknowledgments}\noindent

I thank Professor Peter Hogan for many helpful discussions in the
course of this work and IRCSET and Enterprise Ireland for
financial support.

\vskip 8truepc

\appendix

\setcounter{equation}{0}

\section{Notation and Basic Equations}

Some notation and tensor quantities associated with the vector
field $u^i$ in Section 2 are required. Covariant differentiation
is indicated with a semi--colon and covariant differentiation in
the direction of $u^i$ is denoted by a dot. As usual square
brackets denote skew--symmetrization and round brackets denote
symmetrization. Thus the 4--acceleration of the time--like
congruence is

\begin{equation}\label{A1}
\dot{u}^i:=u^i{}_{;\,j}\,u^j\ .
\end{equation}With respect to $u^i$ and using the projection tensor

\begin{equation}\label{A2}
h_{ij}:=g_{ij}+u_i\,u_j\ ,
\end{equation}the covariant derivative of the 4--velocity $u_{i;\,j}$ can be
decomposed into

\begin{equation}\label{A3}
u_{i;\,j}=\omega_{ij}+\sigma_{ij}+\frac{1}{3}\theta\,h_{ij}-\dot{u}_i\,u_j\
,
\end{equation}with

\begin{equation}\label{A4}
\omega_{ij}:=u_{[i;\,j]}+\dot{u}_{[i}u_{j]}\ ,
\end{equation}the vorticity tensor of the congruence,

\begin{equation}\label{A5}
\sigma_{ij}:=u_{(i;\,j)}+\dot{u}_{(i}u_{j)}-\frac{1}{3}\,\theta\,h_{ij}\
,
\end{equation}the shear tensor of the congruence and

\begin{equation}\label{A6}
\theta:=u^i{}_{;\,i}\ ,
\end{equation}the expansion (or contraction) of the universe.

The Riemann curvature tensor $R_{ijkl}$ is defined by the Ricci
identities

\begin{equation}\label{A7}
u_{i;\,lk}-u_{i;\,kl}=R_{ijkl}\,u^j\ ,
\end{equation}and Einstein's field equations take the form

\begin{equation}\label{A8}
R_{ij}-\frac{1}{2}g_{ij}\,R=T_{ij}\ ,
\end{equation}where $R_{ij}:=R^k{}_{ikj}$ are the components of the Ricci
tensor, $R:=R^i{}_i$ is the Ricci scalar and $T_{ij}$ is the
symmetric energy--momentum--stress tensor. We note that the
coupling constant has been absorbed into $T_{ij}$. With respect to
the 4--velocity field $u^i$ the energy--momentum--stress tensor
can be decomposed as

\begin{equation}\label{A9}
T^{ij}=\mu_0\,u^i\,u^j+p\,h^{ij}+q^i\,u^j+q^j\,u^i+\pi^{ij}\ ,
\end{equation}with

\begin{equation}\label{A10}
q^i\,u_i=0\ , \qquad \pi^{ij}\,u_j=0\ , \qquad \pi^{i}{}_{i}=0\ ,
\end{equation}and $\pi^{ij}=\pi^{ji}$. Then $\mu_0$ is interpreted as the energy
density measured by an observer with 4--velocity $u^i$, $q^i$ is
the energy flow (such as heat flow) measured by this observer, $p$
is the isotropic pressure and $\pi^{i}{}_{j}$ is the trace--free
anisotropic stress (due for example to viscosity).

With respect to $u^i$ the Weyl tensor may be decomposed into its
``electric" and ``magnetic" components given respectively by

\begin{equation}\label{A12}
E_{ij}=C_{ikjl}\,u^k\,u^l\ , \qquad
H_{ij}={}^{*}C_{ikjl}\,u^k\,u^l\ ,
\end{equation}

where ${}^{*}C_{ikjl}=\frac{1}{2}\,\eta_{ik}{}^{mn}\,C_{mnjl}$ is
the dual of the Weyl tensor (the left and right duals being
equal), $\eta_{ijkl}=\sqrt{-g}\,\epsilon_{ijkl}$ with
$g={\rm{det}}(g_{ij})$ and $\epsilon_{ijkl}$ is the Levi-Civita
permutation symbol. The explicit expression for the Weyl tensor in
terms of $E_{ij}$ and $H_{ij}$ is not needed here but can be found
in \cite{Ellis}.

\end{document}